\documentstyle[twocolumn,aps,graphicx,floats]{revtex}

\begin{document}
\draft
 
\wideabs{   
\title{The Mesostructure of Polymer Collapse and Fractal Smoothing}
\author{G. E. Crooks}
\address{Dept. of Chemistry, University of California, Berkeley, CA}

\author{B. Ostrovsky}
\address{Sun Microsystems Computer Company, Chelmsford, MA}

\author{Y. Bar-Yam}
\address{New England Complex Systems Institute, Cambridge, MA}

\date{\today}
\maketitle

\begin{abstract}               
	We investigate the internal structure of a polymer during collapse 
	from an expanded coil to a compact globule.  Collapse is more probable 
	in local regions of high curvature, so a smoothing of the fractal 
	polymer structure occurs that proceeds systematically from the 
	shortest to the longest length scales.  A proposed universal scaling 
	relationship is tested by comparison with Monte Carlo simulations.  We 
	speculate that the universal form applies to various fractal systems 
	with local processes that promote smoothness over time.  The results 
	complement earlier work showing that on the macroscale polymer 
	collapse proceeds by driven diffusion of the polymer 
	ends.
\end{abstract} 

\pacs{87.15.Da, 61.43.-j, 64.60.Cn}
}	

Understanding the collapse of homopolymers from a flexible coil to a 
compact globule is a first step towards modeling the kinetics of 
molecular self-organization.  It may be relevant to a description of 
DNA aggregation and the initial collapse of proteins from an expanded 
state to a molten globule from which the final ordered structure is 
formed.[1,2] We have performed scaling analysis and simulation of this 
transition to investigate kinetic effects during collapse.[3,4] The 
results suggest that the motion of the polymer ends plays an important 
role in kinetics because their motion is constrained only by a single 
bond.  Along the contour monomers have two bonds, their motion is more 
constrained, and aggregation is more difficult.  Thus, collapse for a 
long polymer occurs almost as a one dimensional process where the 
polymer ends accumulate mass by moving along the contour of the 
polymer while accreting monomers and small aggregates.  Encounters 
between monomers far apart along the contour to form rings are rare so 
they play no role in the collapse.  As a result of the faster 
aggregation at the polymer ends the collapsing polymer on a 
macroscopic scale takes on a dumbbell like appearance.  The few DNA 
fluorescence measurements that follow a single polymer collapse and 
its metastable states also indicate the special role of polymer 
ends.[5-7] However, a description of the internal structure of the 
polymer away from the polymer ends has not, thus far, been obtained.

In this manuscript we consider the internal structure of the polymer 
during collapse, not including the ends.  Our objective is to 
understand the local contour structure that consists of small 
aggregates and polymer segments between them.  Our arguments 
generalize the consideration of the freedom of motion of monomers, 
because monomers found in straight segments are much more constrained 
in their motion than monomers in curved segments.  This results in 
faster collapse in regions of high polymer curvature.

We will focus on intermediate length scales between the size of the 
expanded polymer and the size of the collapsed aggregate.  The length 
and time scales to which this analysis is relevant are between the 
size of the initial coil, which scales as $N^{\nu}$ where $N$ is the number of 
monomers and $\nu=0.6$ is the Flory exponent, and the size of the final 
aggregate, which scales as $N^{1/3 }$ (assuming a compact aggregate).  For 
long polymers these scales are well separated.  During collapse, at 
these intermediate length scales, the internal structure of the final 
aggregate as well as of intermediate clusters that are formed should 
not be relevant.  When convenient we can treat clusters as point 
objects, though this is not always necessary.  The dynamic properties 
of cluster movement follows Stokes' Law--Ñthe diffusion constant of 
clusters decreases slowly with cluster size, $D\sim R^{-1}$, where $R$ is the 
radius of a cluster.  This implies that the dynamics of clusters 
varies smoothly from that of the original monomers, and a universal 
scaling behavior of the polymer during collapse should be found.  By 
focusing on intermediate length scales, our results should be widely 
relevant to polymers with varied properties.  While the eventual 
structure of the collapsed polymer depends in detail on 
monomer-monomer interactions, the separation of lengths scales implies 
that for a long enough polymer with a compact final aggregate, the 
details of these interactions should not be relevant to the kinetics 
of the collapse at early times.

To characterize the collapse it is useful to compare the distance 
between two monomers with the contour length of the polymer connecting 
them.  In conventional scaling the polymer end-to-end distance $R$ is 
expressed as a function of the number of monomers $N$, or the number of 
links in the chain $L=N-1$.  When aggregation occurs, the small 
aggregates that form, appearing like beads on a chain, decrease the 
effective contour length of the polymer.  We can define the effective 
contour length by counting the minimum number of monomer-monomer bonds 
that one must cross in order to travel the polymer from one end to the 
other.  Bonds formed by aggregation allow us to bypass parts of the 
usual polymer contour.  Because we are not interested in the structure 
of aggregates we can neglect the difference between different kinds of 
bonds.  In this way the effective number of links in the chain 
decreases over time.  Thus, in order to study the internal polymer 
structure during collapse we investigate, via scaling arguments and 
simulations, the scaling of the end-to-end distance $r(l,t)$ of internal 
polymer segments as a function of their effective contour length $l$.

The equilibrium structure of the polymer before collapse---in good 
solvent conditions---is a self-avoiding random walk, where $r\sim l^{\nu}$, and 
$\nu=0.$6 in three dimensions.  The contour length is proportional to the 
number of monomers.  During collapse, monomers are constrained from 
aggregating with other monomers by their already existing bonds.  A 
completely straight segment of polymer does not allow aggregation 
because no monomer can move to bond with another monomer.  In 
contrast, highly curved regions are more flexible and monomers in 
these regions may aggregate.  Aggregation in a curved region reduces 
the contour length and the polymer becomes straighter, smoothing the 
rough fractal polymer structure.  We therefore expect that the scaling 
exponent will increase over time.  At long enough times the scaling 
will approach that of a straight line ($r\sim l$).  However, this smoothing 
occurs first at the shortest length scales.  In effect the polymer 
structure becomes consistent with a progressively longer persistence 
length.  Assuming scaling, we anticipate that the polymer end-to-end 
distance for a polymer segment away from the ends of contour length 
$l$ will follow the dynamic scaling formula:

\begin{equation}
	r = l f(t/l^{z})
\end{equation}

\noindent The universal function $f(x)$ is a constant for large values of its argument 
so that $r\sim l$ (long times), and scales as $x^{(1-\nu)/z}$ for small values of its 
argument, so that $r\sim l^{\nu}$ (short times).  The short time regime described 
by Eq.~(1) starts after an initial transient (a very short time 
regime) in which no new bonding has taken place.  During the very 
short time regime the time dependence of the universal scaling 
function does not apply.  The usual scaling of the contour length and 
end-to-end distance persists until just after the very short time 
regime because the bonds that are formed initially do not form large 
rings and thus do not affect the large scale polymer structure.  The 
short time regime begins with the first formation of individual bonds 
and lasts until the characteristic relaxation time of the contour of 
length $l$.  This time---the relaxation time of the contour of length 
$l$ ---is the crossover time between the short and long time regimes which 
follows a scaling law $\tau \sim l^{z}$.  The dynamic exponent $z$ is assumed to be 
consistent with conventional Zimm relaxation, $z=3\nu$ .  Finally, we can also 
rewrite this scaling relation in terms of the number of monomers $n$ in 
a polymer segment.  Since the average mass along the contour is $M\sim n/l$
 and $M$ follows power law scaling[3]  $M \sim t^{s}$--- we substitute 
 $l\sim nt^{-s}$ in Eq.~(1) to obtain $r(n,t)$.

We emphasize that kinetic effects become important for collapse of 
polymers in poor solvent, after equilibration in good solventÑthe 
result of a quench in solvent affinity or temperature below the 
thermodynamic transition at $\Theta$-solvent conditions.  Close to the 
$\Theta$-point a mean field argument where kinetics do not play a significant 
role is likely to apply.[8,3] In contrast, we will approximate the 
collapse by a completely irreversible model where no disaggregation 
occurs.  Because the scaling variable that determines the effective 
distance from the $\Theta$-point is  $N^{1/2}\Delta T$, long lengths are equivalent to small 
temperatures, and microscopic reversibility becomes irrelevant at long 
enough length scales.  We are thus consistently adopting a description 
that is valid for lengths longer than the microscopic regime.  We 
further restrict our study to diffusive monomer motion and short-range 
interactions.

\begin{figure}[t]
	\centering
	\includegraphics[scale=0.5]{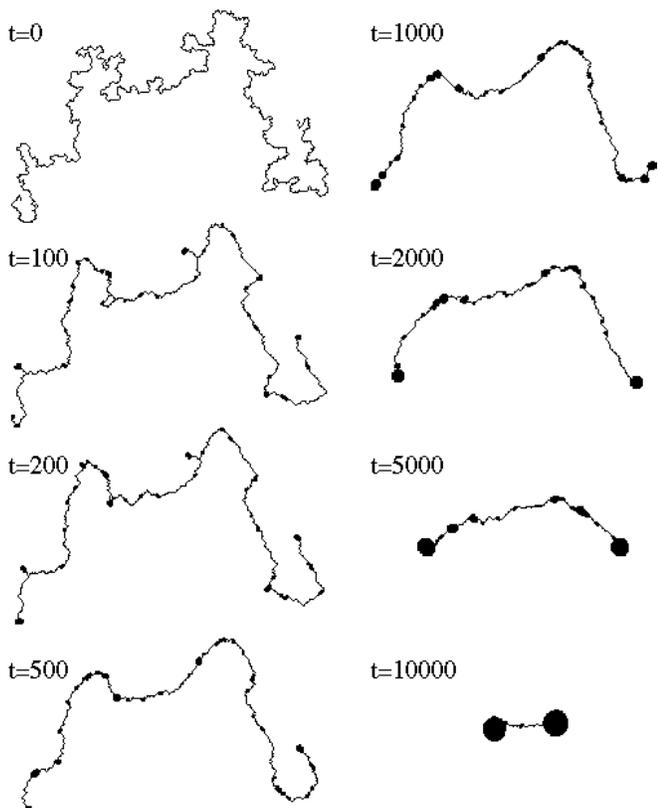} 

 \caption{ 'Snapshots' of the collapse of a single homo-polymer of 
 length $N=500$ monomers in 2-d using the two-space algorithm.  The plot 
 is constructed by placing dots of area $M^{1/2}$ for an aggregate of mass 
 $M$. This does not reflect the excluded volume of the aggregates, which 
 is zero during this collapse simulation.  The frames demonstrate the 
 process of local smoothing that occurs progressively from short to 
 long length scales.  The behavior of the ends is discussed in Refs.~3 
 and 4.  The primary effects of hydrodynamics are included in the 
 simulations by applying Stokes' law to the diffusion of aggregates.  
 }
 \end{figure}

The scaling relationship, Eq.~(1), was tested by Monte Carlo 
simulations.  These simulations in part include the effects of 
hydrodynamics during collapse by scaling the diffusion constant of 
aggregates according to Stokes' Law.  The simulations are based on the 
two-space lattice Monte Carlo algorithm[3,4,9,10] developed for 
simulating high-molecular-weight polymers, and shown to be 
significantly faster than previous state-of-the-art techniques.[9,10] In 
the two-space algorithm odd monomers and even monomers of a polymer 
are distinct and may most easily be described as residing in two 
separate spaces.  Each monomer occupies one cell of a square lattice.  
Both connectivity of the polymer and excluded volume are imposed by 
requiring that, in the opposite space, only the nearest neighbors 
along the contour reside in the $3\times 3\times 3$ neighborhood of cells around 
each monomer.  Motion of monomers is performed by Monte-Carlo steps 
that satisfy the polymer constraints.  Since adjacent monomers (and 
only adjacent monomers) may lie on-top of each other, the local motion 
of the polymer is flexible.  Despite the unusual local polymer 
properties the behavior of long polymers is found to agree with 
conventional scaling results.

The polymer is initially relaxed into an equilibrium configuration 
using a fast non-local ``reptation'' Monte Carlo algorithm.  Monomers 
are randomly moved from one end of the polymer to the other, which, 
for equilibrium geometries, provides equivalent results to the local 
two-space dynamics.

Collapse of the polymer is then simulated using local diffusive 
Monte Carlo dynamics, but without the excluded volume constraint.  
Simulations of a variety of models indicate that excluded volume does 
not significantly affect the kinetics of collapse.[4] Monomers are no 
longer stopped from entering the neighborhoods of other monomers; they 
continue to be required not to leave any neighbors behind.  This 
allows monomers in the same space (odd or even) to move on top of each 
other, and thus aggregate.  Aggregates of any mass occupy only a 
single lattice site, and are moved as a unit by the same dynamics used 
for monomers.  The mass of an aggregate is set equal to the number of 
monomers located at that site.  The probability of moving an aggregate 
is adjusted to be consistent with a diffusion constant that scales by 
Stokes' law for spherical bodies in 3-dimensions, $D \sim M^{-1/3}$.
This represents the effects of hydrodynamics on individual aggregates, but does not 
include coupling of motion of different aggregates.  One time interval 
consists of attempting a number of aggregate moves equal to the number 
of remaining aggregates.

The end-to-end distances of polymer segments, $r$, were measured as a 
function of the effective contour length, $l$.  The effective contour 
length is the minimum number of links along the polymer that connect a 
monomer at one end of the segment with the other end of the segment.  
Since an aggregate occupies only a single lattice site, interior bonds 
of the aggregate need not be counted, and it can be treated like a 
single monomer.  The end-to-end distance is not exactly the Euclidean 
distance between the ends.  Instead it is correctly defined as the 
minimum number of links needed to connect the two ends by any curve in 
space.  Due to the underlying lattice in our algorithm a Manhattan 
metric, with the inclusion of diagonals, is appropriate.

In addition to simulations of polymer collapse in three dimensions we 
also performed simulations of polymers whose motion is confined to two 
dimensions, which are convenient for pictorial illustration (Fig.~1).  
These are not conventional two dimensional simulations because of the 
well known problems with hydrodynamics in two dimensions.[11] Instead, 
they represent the dynamics of a polymer confined at an interface 
(e.g. between two fluids).  Thus the polymer is confined to two 
dimensions while the hydrodynamics is three dimensional.  In this case 
we have  $\nu=0.75$, $z=3\nu$  and $D\sim M^{-1/2}$.  The frames in Fig.~1 illustrate contour smoothing.  
Starting with short length scales, the polymer becomes progressively 
smoother and approaches a straight line.

\begin{figure}
 \begin{center}
	\includegraphics[scale=0.5]{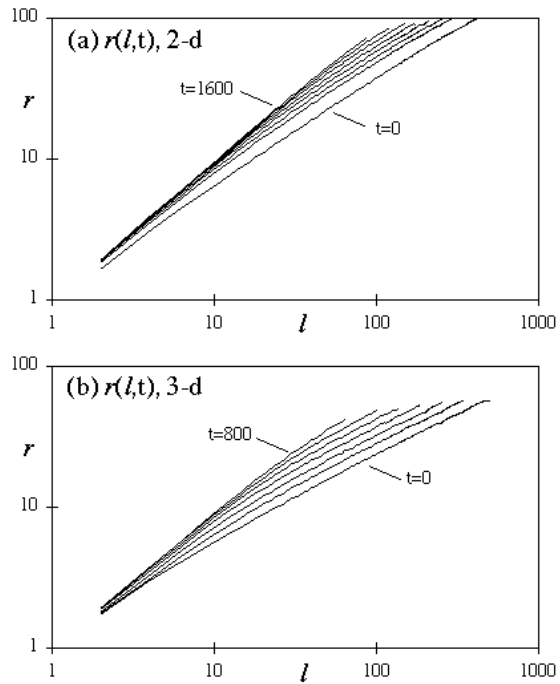} 
\end{center}
 \caption{Plot of the internal polymer segment end-to-end distance $r$, 
 as a function of segment contour length $l$, and collapse time, $t$, in 
 both (a) two dimensions (2-d), for $t=$0, 25, 50, 100, 200, 400, 800, 
 1600 and (b) three dimensions (3-d), with $t=$0, 25, 50, 100, 200, 400, 
 800.  In both cases the polymer contained 500 monomers and results 
 were averaged over 200 collapses.  }
 \end{figure}

In Fig.~2 we show log-log plots of contour length ($l$) versus 
end-to-end distance ($r$) for both 2-d and 3-d simulations.  Initially 
the results are consistent with $r\sim l^{\nu}$ for a self avoiding random walk.  As 
time progresses the polymer becomes smooth resulting in a slope that 
approaches $1$.  The asymptotic behavior can be seen to occur earlier at 
the shortest length scales.

Fig.~3 shows the derivative, obtained from finite differences, as a 
function of time for different segment contour lengths.  For all 
segment lengths the derivative starts at approximately $\nu$ and 
approaches $1$ as the collapse proceeds.  The rate of collapse becomes 
progressively slower as $l$ increases.  The scaling relation, Eq.~(1), 
predicts that the relaxation time will scale with $l$ as $l^{z}$.  Fig.~4 
shows the data following rescaling.  $r/l$ is plotted against the 
rescaled time, $t/l^{z}$.  The generally good coincidence of the different 
curves confirms that the simulation obeys Eq.~(1).[12] An attempt to 
use an asymptotic scaling exponent, $r\sim l^{u}$ , with $u=0.95$, led to a visibly poorer 
fit, as did small variations in the exponent $z$.

The excellent agreement with the expected universal scaling 
relationship using the hydrodynamic exponent, $z=3\nu$ , may be fortuitous 
because the simulations do not contain the full effects of 
hydrodynamics.  Specifically they contain only the effect of 
hydrodynamics on individual aggregates and not the coupling between 
aggregate motion.

 \begin{figure}
 \begin{center}
	\includegraphics[scale=0.5]{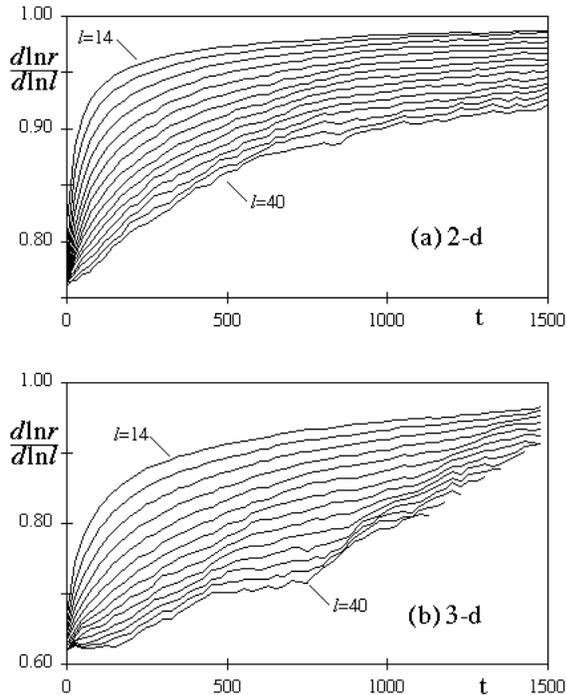} 
\end{center}
 \caption{ Plot of the scaling exponent $d \ln r / d \ln l$ as a function of time for 
 different segment contour lengths $l$.  This figure was obtained from 
 Fig.~2 by calculating average slopes over a segment length of 20.  }
 \end{figure}

In summary, we have found that polymer collapse displays a process of 
fractal smoothing that occurs first at the shortest length scales.  
Our simulations were found to be in good agreement with a universal 
scaling relationship.  It is interesting to speculate that this may 
also apply to other fractal systems where local smoothing processes 
occur.  Several groups have attempted to measure the self-affine 
scaling of horizontal transects of mountain ranges.[13,14] They found 
that a unique fractal dimension, $\mathrm D_{H}$ cannot be assigned, but that the 
effective fractal dimension decreases with length scale.  For example 
Dietler and Zhang [13] have performed calculations for Switzerland, an 
area of $7\times10^{4}\mathrm \, km^{2}$, with a resolution of 100 m.  They 
obtained $\mathrm D_{H}\approx 1.43$ 
at length scales below approximately 5 km, and $\mathrm D_{H}\approx 1.73$ for larger 
length scales.  The data points could also lie on a continuous curve 
rather than two distinct scaling regimes.  Thus the landscape appears 
smoother at shorter length scales.  Short range smoothing may arise 
from processes, such as weathering, that also give rise to short range 
correlations.  Various fractal biological systems formed as a result 
of an initial developmental process may also suffer smoothing as part 
of aging.

\begin{figure} 
 \begin{center}
	\includegraphics[scale=0.5]{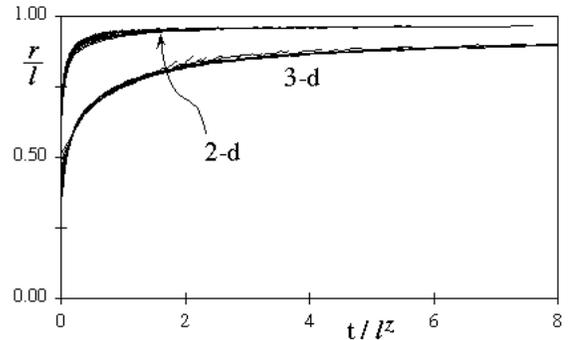} 
\end{center}
 \caption{Plot of the rescaled end-to-end polymer segment distance, 
 $r/l$, as a function of the rescaled time, $t/l^{z}$.  Data for both 2-d and 3-d are 
 shown.  The coincidence of the curves is consistent with validity of 
 the universal scaling relationship, Eq.~(1).  }
 \end{figure}

Since this work[15] was completed, a number of other works have explored 
the kinetics of collapse using simulations, analytic treatments and 
scaling arguments.  Timoshenko, Kuznetsov and Dawson[16] studied the 
kinetics of collapse using Monte Carlo simulations and a mean-field 
``Gaussian self-consistent'' approach.  Their Monte Carlo simulations 
are based upon an underlying lattice model which is similar to ours, 
however, they do not move aggregates as a unit.  Since they only move 
individual monomers, Monte-Carlo rejection of moves causes the 
diffusion constant of aggregates to decreases very rapidly with 
aggregate size (naively, it decreases exponentially [17]).  By contrast, 
in a fluid, collective motion results in Stokes' law diffusion, which 
we have included in our simulations.  The slow diffusion of aggregates 
in their simulations cause their results to be distinct from ours.  
From their figures it appears clear that aggregates tend to pin the 
polymer contour.  Their Gaussian self-consistent approach is 
analytically elaborate, however, it is not clear from their analysis 
whether it treats correctly the diffusion of clusters.  Moreover, 
since some equilibrium scaling laws are not correct in this method it 
is hard to evaluate whether the kinetic properties are correct and 
their analysis does not clarify this point aside from the claim that 
the analytic results are in aggrement with their Monte Carlo 
simulations.

Buguin, Brochard-Wyart and de Gennes[17] have presented scaling arguments 
based on a model of local clusters ``pearls'' forming during collapse 
close to the $\Theta$-point in the mean field regime where surface tension is 
the driving force of collapse.  Pitard[18] has further considered the 
dynamics of collapse in this mean field regime by discussing the 
effect of tension along a polymer contour between two clusters 
(pearls) and extended the arguments to considerations of a string of 
clusters.  These papers refer to a different regime (i.e.  the mean 
field regime) than our analysis.  Within this regime they provide 
complementary insights about the structure of clusters or pearls 
during collapse and the formation of a globule, which is important 
both to the kinetics of collapse and to the eventual structure of the 
aggregate that is formed at the end.  It is worth noting that our 
simulations do not allow monomer motion along the polymer contour 
which can allow monomers to leave and join aggregates.  The 
distribution of cluster sizes may be affected by such motion.  We 
note, however, that the essential results of this paper should not be 
changed by redistribution of monomers along the contour, and resulting 
change in the distribution of the sizes of clusters, because they only 
affect the distribution of diffusion constants which vary only weakly 
with aggregate size.  Moreover, the scaling law Eq.~(1) does not 
refer to aggregate size and should not be affected.

Finally, Kantor and Kardar[19] have investigated the properties of 
charged polymers and find their compact form exhibits a necklace shape 
with end aggregates and intermediate aggregates forming as a function 
of the charge density.  These results also display some interesting 
similarities to collapse behavior and further research may reveal a 
connection between their results and the studies of collapse.

We would like to thank A. Grosberg and M. Kardar for helpful 
discussions.  A referee is to be acknowledged for pointing out that 
the universal scaling law does not apply in the limit $t\rightarrow 0$ before the first 
bonding events.




\begin{references}
\bibitem{12} In the limit $t\rightarrow 0$ the scaling relation 
predicts $r/l\rightarrow \infty$ because a backwards extrapolation of 
the collapse allows arbitrarily small persistance lengths and 
infinitely many monomers ($n \rightarrow \infty$).  The finite values seen 
in Fig.~4 at $t\rightarrow 0$ are due to finite size scaling effects at the singular point $t=0$.
\end{references}
\end{document}